# Using Deep Learning to Identify Initial Error Sensitivity for Interpretable ENSO Forecasts

Kinya Toride,[a,b] Matthew Newman,[a] Andrew Hoell,[a] Antonietta Capotondi,[a,b] Jakob Schlör,[c] Dillon J. Amaya,[a]

[a] *Physical Sciences Laboratory, National Oceanic and Atmospheric Administration, Boulder, Colorado*

[b] *Cooperative Institute for Research in Environmental Sciences, University of Colorado Boulder, Boulder, Colorado*

[c] *European Centre for Medium Range Weather Forecasts (ECMWF), Reading, UK*

*Corresponding author*: Kinya Toride, kinya.toride@noaa.gov

ABSTRACT

We introduce an interpretable-by-design method, optimized model-analog, that integrates deep learning with model-analog forecasting which generates forecasts from similar initial climate states in a repository of model simulations. This hybrid framework employs a convolutional neural network to estimate state-dependent weights to identify initial analog states that lead to shadowing target trajectories. The advantage of our method lies in its inherent interpretability, offering insights into initial-error-sensitive regions through estimated weights and the ability to trace the physically-based evolution of the system through analog forecasting. We evaluate our approach using the Community Earth System Model Version 2 Large Ensemble to forecast the El Niño–Southern Oscillation (ENSO) on a seasonal-to-annual time scale. Results show a 10% improvement in forecasting equatorial Pacific sea surface temperature anomalies at 9–12 months leads compared to the unweighted model-analog technique. Furthermore, our model demonstrates improvements in boreal winter and spring initialization when evaluated against a reanalysis dataset. Our approach reveals state-dependent regional sensitivity linked to various seasonally varying physical processes, including the Pacific Meridional Modes, equatorial recharge oscillator, and stochastic wind forcing. Additionally, forecasts of El Niño and La Niña are sensitive to different initial states: El Niño forecasts are more sensitive to initial error in tropical Pacific sea surface temperature in boreal winter, while La Niña forecasts are more sensitive to initial error in tropical Pacific zonal wind stress in boreal summer. This approach has broad implications for forecasting diverse climate phenomena, including regional temperature and precipitation, which are challenging for the model-analog approach alone.

SIGNIFICANCE STATEMENT

This study demonstrates that combining deep learning and a simple analog forecasting method can yield skillful and interpretable El Niño–Southern Oscillation forecasts. A convolutional neural network is used to find critical areas for picking analog members. This is important because it is challenging to explain the decision-making processes of recent deep-learning approaches. The developed approach can be applied to various climate predictions.

# 1. Introduction

The prediction of climate variability over seasonal to interannual time scales greatly depends on the quality of El Niño–Southern Oscillation (ENSO) forecasts. The magnitude and pattern of tropical sea surface temperature (SST) anomalies associated with ENSO influence global climate through atmospheric teleconnections primarily driven by the Walker and Hadley circulations and stationary Rossby wave trains (Alexander et al. 2002; Hoell and Funk 2013; Capotondi et al. 2015; Taschetto et al. 2020). However, state-of-the-art atmosphere-ocean coupled models do not exhibit a substantial improvement over simpler linear models in predicting ENSO (Newman and Sardeshmukh 2017; Shin et al. 2021; Risbey et al. 2021).

With recent progress in deep learning, several studies have applied various neural networks to ENSO prediction (Ham et al. 2019; Petersik and Dijkstra 2020; Cachay et al. 2021; Chen et al. 2021; Ham et al. 2021; Zhou and Zhang 2023). Considering the data-intensive nature of deep learning, long-term climate simulations from multiple models are often leveraged to capture nonlinear dynamics of ENSO and mitigate model-specific biases. While these data-driven models exhibit promising performance, interpreting their decision-making processes poses a challenge due to the large number of hidden parameters. The interpretability of prediction models is crucial since models with better interpretability can enhance scientific understanding of physical processes, which can, in turn, improve prediction skill. Explainable artificial intelligence (XAI) is frequently used to elucidate neural network models in a post-hoc manner (e.g., Shin et al. 2022). However, different XAI techniques may yield different explanations for the same deep learning model (Mamalakis et al. 2022), and it remains challenging to explain complex models despite their superior accuracy in general.

Analog forecasting is a simpler method which makes predictions based on similar states that occurred in the past, assuming they follow the attractor of the dynamical system (Lorenz 1969a). While the sample size of historical records is too small to find good analogs for most climate-scale applications (Van den Dool 1989), simulated climate data allow for drawing "model-analogs" from thousands of years of data (Ding et al. 2018). Analog forecasting circumvents issues with initialization shock—rapid adjustment processes due to an imbalance between initial conditions and model dynamics (Mulholland et al. 2015)—and model drift—the development of forecast errors over time due to model biases (Magnusson et al. 2013). By directly using trajectories on the model's own attractor, this method provides comparable skill to that of coupled atmosphere-ocean models in forecasting seasonal tropical SST (Ding et al. 2018, 2019).

However, despite advances, finding reliable analogs within the chaotic climate system remains challenging due to both the limited sample size, even with thousands of years, and model imperfections leading to disparities between the model attractor and nature attractor. In chaotic systems, even tiny disturbances in initial states can lead to significantly divergent trajectories (Lorenz 1963, 1969b). Fig. 1b illustrates this issue, showing that a few model-analogs, selected based only on minimal mean-square differences across the tropics, can evolve into the opposite phase of ENSO within 12 months.

Alternatively, there may exist trajectories with slightly different initial conditions that remain closer to the true trajectory over some period of time (Grebogi et al. 1990; Judd et al. 2004). Identifying these shadowing trajectories involves considering the sensitivity to initial conditions, with certain regions being more sensitive to initial errors while others are relatively insensitive (Errico 1997; Barsugli and Sardeshmukh 2002). For instance, the North Pacific Meridional Mode (NPMM) serves as one of key ENSO precursors (Chiang and Vimont 2004; Amaya 2019), driving the search for analogs that closely match over the NPMM region. Essentially, we aim to assign higher weights to initial-error-sensitive regions, thereby optimizing the selection of model-analogs so that their subsequent trajectories will more closely shadow the true trajectory.

In this study, we introduce a deep learning hybrid method, where a convolutional neural network predicts state-dependent weights for selecting "optimized model-analogs". The combination of analog forecasting and machine learning has been investigated by several studies. Chattopadhyay et al. (2020) clustered surface temperature patterns into five groups and used a capsule neural network to predict the cluster indices based on states 1–5 days prior. Rader and Barnes (2023) introduced the idea of training a neural network to learn weights of a global mask to improve the selection of model-analogs for analog forecasting, and then used their mask to explore sources of predictability. However, their approach is state-independent and their forecasts struggle to predict extreme events.

Here, we find a pattern of weights identifying where the model-analogs should most closely match each initial (target) anomalous state. That is, regions with higher weights are those where initial errors may have a greater impact on subsequent anomaly evolution. Fig.

1c illustrates that optimized model-analogs selected using predicted weights exhibit smaller error growth compared to the original model-analogs.

Our forecasting method is an interpretable-by-design approach, blending deep learning with interpretable methods (Chen et al. 2019; Rudin 2019). We decompose the forecasting processes into two components: determining the best initial state matches and tracking subsequent evolution through the analog method. Specifically, this approach offers two key advantages in terms of interpretability. First, the estimated weights show regions where error growth is particularly sensitive to initial condition error. These weights, which serve as the network's explanations or reasoning processes, are directly used for analog forecasting and integrated in the training process (ante-hoc). In contrast, XAI methods offer post-hoc explanations, which are not the actual reasoning used in decision processes. Second, once analogs are identified using weights, we can trace the physically-based evolution of any other field available in the model simulation for any lead time. This is a key advantage of the model-analog technique that is unattainable with a standalone neural network unless it is trained for all variables.

Our approach improves forecast skill of equatorial Pacific SST in both perfect-model and hindcast experiments. While many machine learning-driven studies typically focus on predicting simple Niño indices (Ham et al. 2019; Petersik and Dijkstra 2020; Cachay et al. 2021; Chen et al. 2021; Ham et al. 2021; Shin et al. 2022), we aim to improve the prediction of the spatial pattern of equatorial Pacific SST given the considerable diversity of individual ENSO events (Capotondi et al. 2015). We describe our data and methods in Section 2, then evaluate forecast skill in perfect-model experiments in Section 3. In Section 4, we demonstrate the connection between the predicted weights and various physical processes associated with ENSO dynamics, including the asymmetry in initial-error-sensitivity for El Niño and La Niña. Section 5 presents the application of the developed method to hindcast experiments. Finally, Section 6 summaries our results.

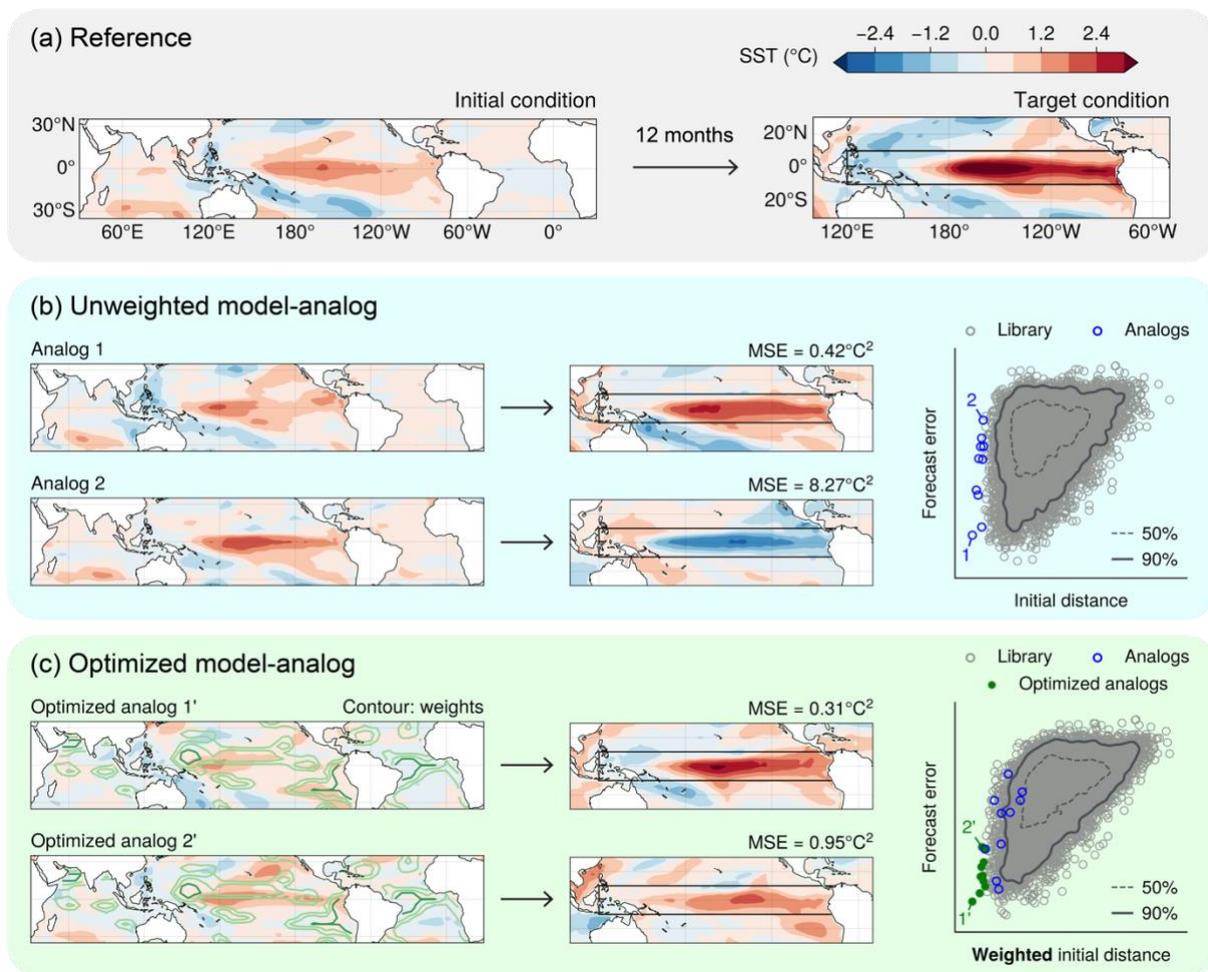

Fig. 1. Schematic method overview of the current study. (a) Reference initial condition for analog selection and target condition 12 months after. The black box in the target condition represents the equatorial Pacific, which is the focus area in this study. (b) Unweighted model-analogs and corresponding forecasts for the best and worst analogs. The mean square errors (MSEs) of the forecasts are shown in each panel. The scatter plot shows initial distances and forecast errors for all samples in the library, along with smoothed probability density curves. Blue circles show 10 analogs with the smallest initial errors. (c) As in (b), but for the optimized model-analogs which exhibit smaller error growth compared to the original analogs. This method uses deep learning to derive optimized weights for analog selection, displayed by contour lines. The scatter plot uses weighted initial distances on the x-axis. Green circles represent 10 optimized analogs, which may be compared to the original analogs represented by blue circles.

## 2. Methods

*a. Data*

We first evaluate the hybrid deep learning and model-analog approach within a perfect-model framework, with the same climate model generating training, validation, and test datasets. We use an ensemble of historical simulations from the Community Earth System Model Version 2 Large Ensemble (CESM2-LE; Rodgers et al. 2021). The CESM2-LE historical simulation consists of 100 ensemble members during 1850–2014, resulting in 16,500 years of data. We use monthly mean sea surface temperature (SST), sea surface height (SSH), and zonal wind stress (TAUX) data. These data are interpolated to two different resolutions, $2° \times 2°$ and $5° \times 5°$. The coarser resolution data are used to train the neural network model and to select analogs, while the finer resolution data are used as forecasts after finding analogs. Detrended anomalies are determined by removing the ensemble mean temporally smoothed with a 30-year centered running mean at each grid point. Throughout this study, we exclusively use anomalies. We partition the dataset into training (1865–1958; 9400 years, 70%), validation (1959–1985; 2700 years, 20%), and test (1986–1998; 1300 years, 10%) subsets. The training dataset is also used as the library to select model-analogs.

To test the trained model with observed estimates, we use the Ocean Reanalysis System 5 (ORAS5; Zuo et al. 2019) interpolated to the fine and coarse resolution grids. This evaluation uses a fair-sliding anomaly approach that refrains from using future data not available at the time of the forecast (Risbey et al. 2021). Specifically, anomalies are determined by removing the mean and linear trend during the prior 30 years up to the year of the current forecast. Note that our model is not trained on any reanalysis data.

*b. Architecture of the optimized model-analog approach*

We develop a deep learning method to predict weights based on a specified initial condition. To reduce computational cost, we use the coarse resolution data over 50°S–50°N (13 latitudes × 72 longitudes × 3 variables) as our input. The architecture of the optimized model-analog approach is depicted in Fig. 2. Our chosen model is the U-Net (Ronneberger et al. 2015), a fully convolutional network consisting of a symmetrically designed downsampling encoder followed by an upsampling decoder. We also experimented with variations such as U-Net with residual blocks (He et al. 2015) and with attention gates (Oktay et al. 2018), but found minimal differences.

The encoder in our architecture consists of stacked blocks, each including two convolutional layers and a max pooling operation, halving the spatial resolution while doubling the channel size (i.e., last dimension). Mirroring the encoder, the decoder includes similar stacked blocks where each incorporates a transposed convolutional layer followed by two convolutional layers. This setup reverses the encoder's blocks by doubling the spatial resolution and reducing the channel size by half. Additionally, skip connections concatenate the features from the downsampling encoder into the decoder at the corresponding level. A final 1×1 convolution aligns the output channel size with the number of input variables.

Two hyperparameters, namely depth and initial channel size, greatly influence the network size. Here, depth corresponds to the number of blocks in the encoder, set at 4 in this study. The initial channel size (indicated by $M$ in Fig. 2) is the output channel size of the first encoder block, set at 256 in our study. Either increasing the depth by one or doubling the initial channel size quadruples U-Net parameters. The sensitivity of the obtained results to the network size is discussed in Text S1.

The U-Net predicts weights that are used to determine weighted initial distances from the input initial condition for every sample within the library. The library comprises all states from the training dataset of the corresponding calendar month, which introduces seasonal cycle effects. The weighted initial distance ($d_i$) between the target state and each library state (sample index $i$) is defined as the sum of weighted mean square errors (wMSE) of standardized SST, SSH, and TAUX anomalies over 50°S–50°N,

$$d_i = \text{wMSE}_i(\text{SST}) + \text{wMSE}_i(\text{SSH}) + \text{wMSE}_i(\text{TAUX}) , \qquad (1)$$

where wMSE of the standardized anomalies is defined as:

$$\text{wMSE}_i = \frac{\sum_j w_j \cos \phi_j \left( \frac{x_j}{\sigma_X} - \frac{y_{i,j}}{\sigma_Y} \right)^2}{\sum_j w_j \cos \phi_j} \qquad (2)$$

Here, $j$ represents a spatial degree of freedom, $w$ represents the weight predicted by U-Net, $\phi$ denotes latitude, $\cos \phi$ accounts for the grid area weight, $x$ represents the input initial state, and $y$ represents a state in the library (sample index $i$). Additionally, $\sigma_X$ and $\sigma_Y$ represent the square root of domain-averaged temporal variance over the input domain, used to standardize units across different variables and correct model biases. Note that for $w_j = 1$, $d$ is essentially the same as the distance metric used by Ding et al. (2018) to determine unweighted model-analogs.

The most intuitive training method might be selecting analogs with the smallest weighted initial distances and defining a loss function based on analog forecast errors. However, this approach involves the complex time evolution of the climate model, with unknown analytical derivatives. Updating network parameters through backpropagation would require calculating the gradient of the climate model with respect to these parameters. While finite difference methods can approximate the gradient of an unknown function, this approach is computationally expensive and may suffer from numerical instability. Given these computational challenges, we opt for a more efficient strategy to update model parameters.

Initially, the weighted initial distances are sorted, and samples with the lowest weighted initial distances are selected, specifically the top 2% of samples (dark blue circles in Fig. 2). We focus on these subsamples so that the network is not affected by samples that significantly deviate in initial conditions. As the network is updated and predicts different weights, a different set of subsamples is selected. Note that the sensitivity to the number of retained samples is relatively low. The loss function is defined as the mean-square-error (MSE) between the normalized weighted initial distances ($d$) and forecast errors ($e$) of the chosen subsamples, where the forecast error is defined as the MSE of SST over the equatorial Pacific (10°S–10°N, 120°E–70°W; black box in Fig. 1) at a certain lead time ($\tau$). Here forecast errors are included in the loss function as the target of the neural network, which do not depend on network parameters. The loss function $L_k$ for the given initial condition (sample index $k$) can be expressed as:

$$L_k = \frac{1}{n_{sub}} \sum_i^{n_{sub}} \left( \frac{d_i}{\max_{i \in n} d_i} - \frac{e_i(\tau)}{\max_{i \in n} e_i(\tau)} \right)^2 \quad (3)$$

where $i$ represents the index of samples, $n_{sub}$ represents the number of subsamples (i.e., 188 in the present study), and $n$ represents the number of samples in the library. The weighted initial distances and forecast errors are scaled by the respective maximums. Minimizing the loss guides the U-Net to optimize weights so that samples with smaller forecast errors have smaller weighted initial distances. Essentially, the objective is to maintain consistency in initial and forecast errors across the subsamples. This iterative process is executed for each sample in the training dataset, constituting one epoch.

Although the U-Net can be trained for various lead times ($\tau$), it then results in identifying different analogs for different lead times. This compromises one of the advantages of analog forecasting: the ability to track the time evolution of the system. To address this, we train the U-Net using forecast errors ($e$) defined by the mean of MSEs across 3, 6, 9, and 12-month lead times over the equatorial Pacific. This approach yields comparable skill to training for specific lead times of 6, 9, or 12 months, as detailed in Text S3. The final loss function is defined as:

$$L_k = \frac{1}{n_{sub}} \sum_{i}^{n_{sub}} \left( \frac{d_i}{\max_{i \in n} d_i} - \frac{e_i}{\max_{i \in n} e_i} \right)^2 \qquad (4)$$

where forecast error is defined as:

$$e_i = \frac{1}{4} \sum_{\tau=3,6,9,12} e_i(\tau) \qquad (5)$$

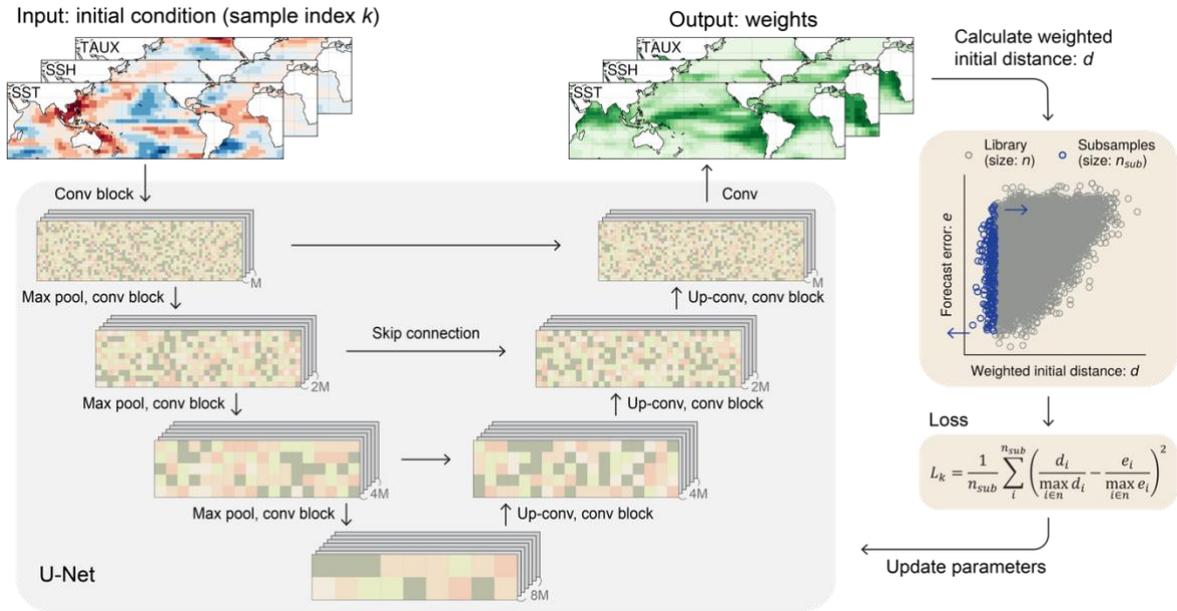

Fig. 2. Architecture of the optimized model-analog approach. The input variables are SST, SSH, TAUX at a 5° × 5° resolution between 50°S–50°N. The U-Net consists of convolution blocks, max pooling layers, transposed convolutional layers, and skip connections. Each convolution block includes two sets of convolutional layers, batch normalization, and ReLU activation. The final 1×1 convolutional layer predicts weights for each variable. $M$ denotes the initial channel size, which is set to be 256 in this study. The predicted weights determine the weighted initial distances for every sample within the library. The top 2% of samples (dark blue circles) are then used to calculate the loss function. This loss function updates the U-Net parameters so that samples with smaller forecast errors have smaller weighted initial distances (indicated by dark blue arrows in the scatter plot).

During each epoch, we monitor ensemble-mean forecast error at 12 months lead. Here, we choose 30 analog members (see Text S2 for details). Note that this member size differs from the subsample size used for calculating the loss function. The loss function is designed to learn the relationship between initial and forecast errors by using a larger sample that includes both good and bad analogs. However, actual forecasts are generated using a smaller sample size to ensure that only the best analogs are selected. While we observe that the forecast error tends to decrease with more analog members once the model is properly trained, we choose to use 30 members to correspond with the unweighted model-analog approach. The maximum number of epochs is capped at 60, and we use early stopping to prevent overfitting, i.e. training is stopped when the ensemble-mean forecast error in the validation dataset ceases to decrease. The Adam optimizer (Kingma and Ba 2017) is used to update network parameters. We train the model 10 times to account for the random initialization of U-Net parameters, and present the average result across these 10 trained models. Since analog selection is performed within the library of the corresponding month, we train a separate U-Net for each month. The source code is available on GitHub (https://github.com/kinyatoride/DLMA).

*c. Hyperparameter tuning*

Key hyperparameters considered in this study are the initial channel size, depth, learning rate, and subsample size. In the initial phase of hyperparameter tuning, we focus on January initialization with a lead time of 12 months. This choice is motivated by the largest ENSO variability observed during this month in the model. All hyperparameters are optimized based on ensemble-mean forecast error in the validation dataset with a 12-month lead time. The learning rate is optimized randomly within the range of 1.0e-6 to 1.0e-4, and the subsample size within 0.3% to 5% of the total sample size. The selected learning rate is 1.5e-5, and the subsample size is 2%. Details regarding hyperparameter tuning related to the network size are discussed in Text S1.

Upon completing the tuning process, the same set of hyperparameters is adopted for other initialization months, except for the learning rate. Due to the significant impact of the learning rate, we fine-tune this parameter independently for each month, ranging from 1.0e-5 to 3.5e-5 depending on the month.

*d. Unweighted model-analog and neural network-only approach*

We compare our hybrid approach against both the original (unweighted) model-analog approach and an equivalent neural network-only approach.

The original model-analog approach draws analogs based on unweighted distance (Ding et al. 2018, 2019; Lou et al. 2023). Here, distance is defined as the sum of MSEs of standardized SST and SSH over 30°S–30°N. MSE is similar to the formulation in Eq. (2) but with a constant weight ($w_j = 1$). The number of analog members is set to 30. In contrast to the hybrid method, distances are calculated using the 2° data since no training is required. TAUX and extratropical regions are omitted in this approach, as their inclusion has been found to degrade skill of the unweighted model-analog approach. More discussion can be found in Text S2.

To address the question of whether combining deep learning and analog forecasting might degrade the deep learning capabilities, we compare with a neural network-only method using a similar architecture. We use the same U-Net architecture except for the final layer. The final 1×1 convolution is adjusted to generate fine-resolution SST fields over the equatorial Pacific. Consequently, this approach takes 5° SST, SSH, and TAUX fields over 50°S–50°N as input and predicts 2° SST over the equatorial Pacific. Given the discrepancy in dimension sizes between inputs and outputs, we apply additional padding and cropping of the data. The number of trainable parameters in this modified U-Net differs from the original by less than 0.01%. While the initial channel size and depth are the same as the original, we tune the learning rate separately for this model. Note that this model is only evaluated for January initialization.

*e. Evaluation of state-dependent weights significance*

We conduct additional three experiments to evaluate the significance of the state-dependent aspect of weights. The first experiment (referred to as the "mean weights" experiment) selects model-analogs using the overall (year-round) mean weights, determined by averaging the weights from all initializations and ensembles in the test dataset. The second experiment (referred to as "seasonal weights") uses the mean weights from the corresponding month initialization, allowing for seasonality in the weights. These two experiments evaluate whether state-independent weights, with and without seasonality, can reproduce the results obtained with state-dependent weights. The last experiment (referred to as "asymmetric weights") tests our system's ability to capture the asymmetry of tropical ocean dynamics. For this, we input opposite sign anomalous initial conditions to the trained model to predict weights. When selecting analogs, we use the original sign initial conditions with the predicted weights.

*f. Evaluation metrics*

We use root-mean-square error (RMSE) and squared (uncentered) anomaly correlation ($AC^2$) to assess the performance of ensemble-mean forecasts. $AC^2$ is specifically defined as $AC^2 = (\max(AC, 0))^2$, ensuring that negative correlations are treated as zero. We use squared anomaly correlation instead of anomaly correlation because it indicates the fraction of the variance described by the model when average anomalies are zero.

To test the statistical significance of the improvements achieved through the optimized analog approach over the unweighted approach, we conduct a one-sided permutation test using the time-series of paired forecasts. For instance, if the forecasts from the optimized approach are represented as $x = [x_1 \ x_2 \ x_3]$ and those from the unweighted approach as $y = [y_1 \ y_2 \ y_3]$, these can be permuted to form a new pairs, such as $x = [y_1 \ x_2 \ y_3]$ and $y = [x_1 \ y_2 \ x_3]$. The null hypothesis is that the true improvement is zero, which is rejected at the significance level of 5%. The null distribution is constructed through 10,000 permutations. When multiple hypotheses are simultaneously tested, as for a map of gridded data, Wilks (2016) recommends adjusting the threshold p-value for the number of false discoveries. We use the Benjamini and Hochberg step-up procedure (Benjamini and Hochberg 1995) with a 5% false discovery rate.

To evaluate the probabilistic skill, we use the continuous ranked probability score (CRPS), which is the integral of the squared difference between cumulative distribution functions and corresponds to the integral of the Brier score over all possible threshold values. CRPS can be decomposed into three components: reliability, resolution, and uncertainty (Hersbach 2000). Reliability (negatively oriented) is related to the flatness of the rank histogram, which is the frequency distribution of the rank of the verification relative to sorted forecast ensembles (Hamill 2001). A flat rank histogram indicates reliable forecasts, while a U-shaped rank histogram suggests underdispersed ensembles, meaning that the verification is often an outlier among the ensembles. Resolution (positively oriented) is linked to the ensemble spread and its outliers, where narrower spread typically implies larger resolution. Uncertainty reflects the inherent variability in the cumulative distribution of the verification, which remains unaffected by the forecasts. Climatological forecasts are by definition perfectly reliable but offer no resolution.

## 3. Forecast verification

*a. January initialization*

Fig. 3 shows perfect model skill using both unweighted and optimized model-analog methods for January initialization, with the test dataset spanning 1,300 years. The application of deep learning significantly enhances analog selection for forecasting SST patterns over the equatorial Pacific. RMSE is reduced by 10% for a lead time of 9–12 months (Fig. 3a), and $AC^2$ of 0.4 is extended by more than 2.5 months (Fig. 3b). These improvements remain robust and are minimally affected by random initialization of the training, as indicated by the orange shade. However, for shorter lead times (i.e., 1–2 months lead), the optimized approach exhibits worse forecast errors, suggesting that the neural network assigns more weights to regions beyond the target area to select analogs with better forecasts in longer leads. Consequently, the unweighted approach, which allocates relatively more weights over the equatorial Pacific, results in lower forecast errors for shorter leads.

Figs. 3c and 3d illustrate the spatial distribution of RMSE reduction and the increase in $AC^2$ achieved by the optimized approach. Skill is consistently improved east of the Maritime Continent, particularly around the Niño 3.4 region in the central equatorial Pacific. However, over the Maritime Continent, neither RMSE nor $AC^2$ exhibits significant improvements, primarily due to the small SST variability in the region and the use of MSE in the loss function. The hybrid approach enhances skill in the central equatorial Pacific, where unweighted model-analogs exhibit the highest skill (Ding et al. 2018).

Although the optimized model-analog approach significantly improves analog forecasting, we might wonder whether a standalone neural network would produce better forecasts. Figs. 3a and 3b also display the forecast skill of the equivalent neural network-only method; importantly, this model was trained separately for 3, 6, 9, and 12 months leads. While the neural network-only method exhibits better skill at 3 and 6 months leads, it demonstrates similar skill at 9 and 12 months leads. With respect to $AC^2$, the optimized model-analog approach shows better accuracy at these leads, where this approach exhibits largest improvements (see Text S3). These results demonstrate that combining neural networks with model-analogs not only improves tracking climate state evolution, but also yields comparable forecast skill compared to a neural network-only approach with a similar architecture and training efforts.
14

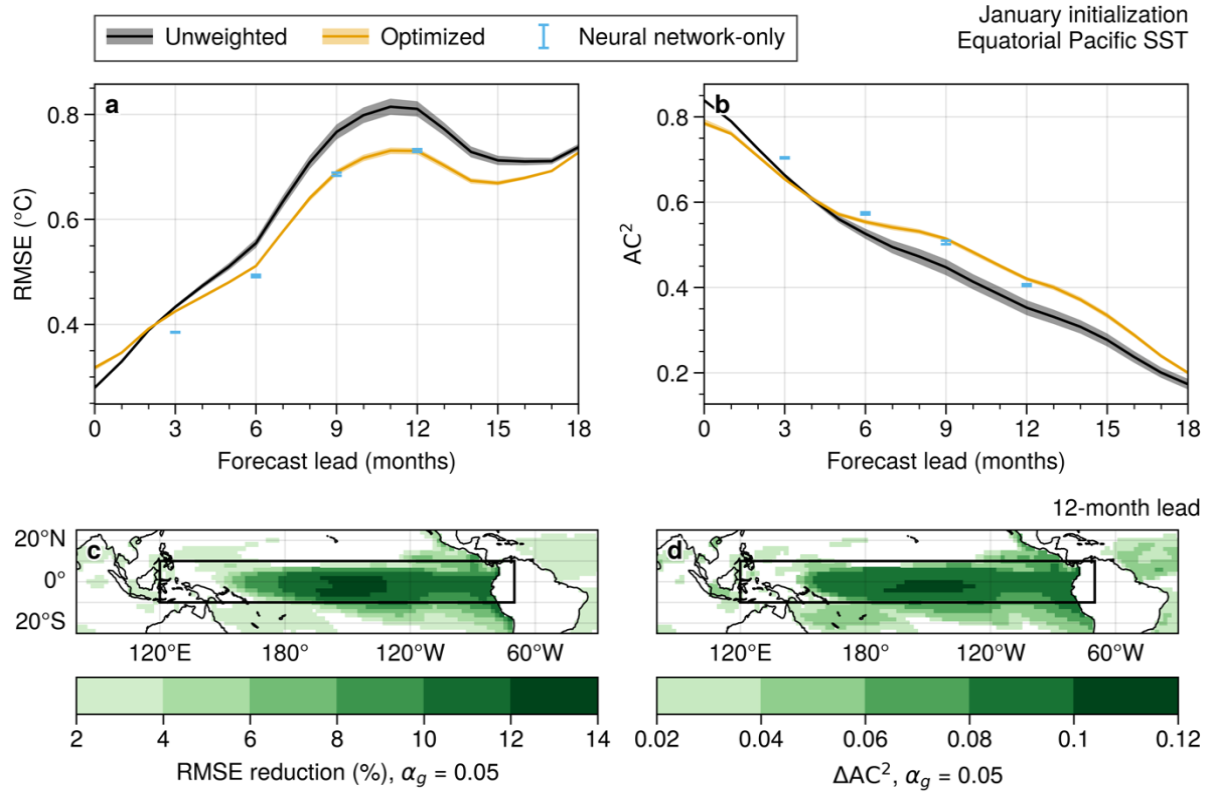

Fig. 3. Forecast skill comparison among the unweighted model-analog, optimized model-analog, and neural network-only approaches for January initialization using the test dataset. (a) Root-mean-square error (RMSE) of equatorial Pacific SST as a function of forecast lead. The black shading represents the 95% confidence interval estimated through the permutation test between unweighted and optimized results. The orange shading and blue error bars show the spread due to random initialization of network parameters. (b) Similar to (a), but for squared anomaly correlation ($AC^2$) averaged over the equatorial Pacific. (c) RMSE reduction (%) of 12-month lead SST by the optimized approach compared to the unweighted approach. (d) Similar to (c), but for the increase in $AC^2$. In (c) and (d), color shading indicates statistically significant improvements at the 5% level adjusted with the 5% false discovery rate.

*b. Seasonal, state-dependent, extreme, and probabilistic skill analysis*

Having tuned the hyperparameters for January initialization, we extend the optimized model-analog approach to other initialization months. Fig. 4 shows the seasonal variation of perfect-model $AC^2$ averaged over the equatorial Pacific. Fig. 4c shows that optimized model-analogs generally yield consistent impacts on analog forecasting across all initialization months. While the forecast skill tends to be reduced for shorter leads, typically ranging from 0 to 3 months, since the neural network places more weights outside the target region, substantial improvements are made for longer leads ranging from 6 to 18 months. These improvements are particularly notable for initialization during boreal winter and spring (Nov–Apr), with verification during boreal fall and winter (Sep–Mar).

To evaluate the contribution of the state-dependent aspect of weights to the observed skill improvements, Figs. 4d–f show the differences in $AC^2$ with modified weights experiments. In general, model-analogs selected with modified weights outperform the unweighted approach. When weights are state-independent and lack seasonality, there is no observed skill reduction for shorter leads, but the improvements at longer leads are less significant compared to the state-dependent approach (Fig. 4d). Fig. 4e indicates that while the seasonality of weights increases skill, the improvements are still not as significant. Fig. 4f shows a reduction in skill improvements when weights are estimated using asymmetric inputs. The skill improvements compared to the state-dependent optimized model-analog approach are approximately 40% for the mean weights, 50% for both the seasonal and asymmetric weights experiments at 9–15 months leads on average. These findings suggest that state-dependent weights are necessary to identify shadowing trajectories at longer leads and thereby enhance forecast skill.

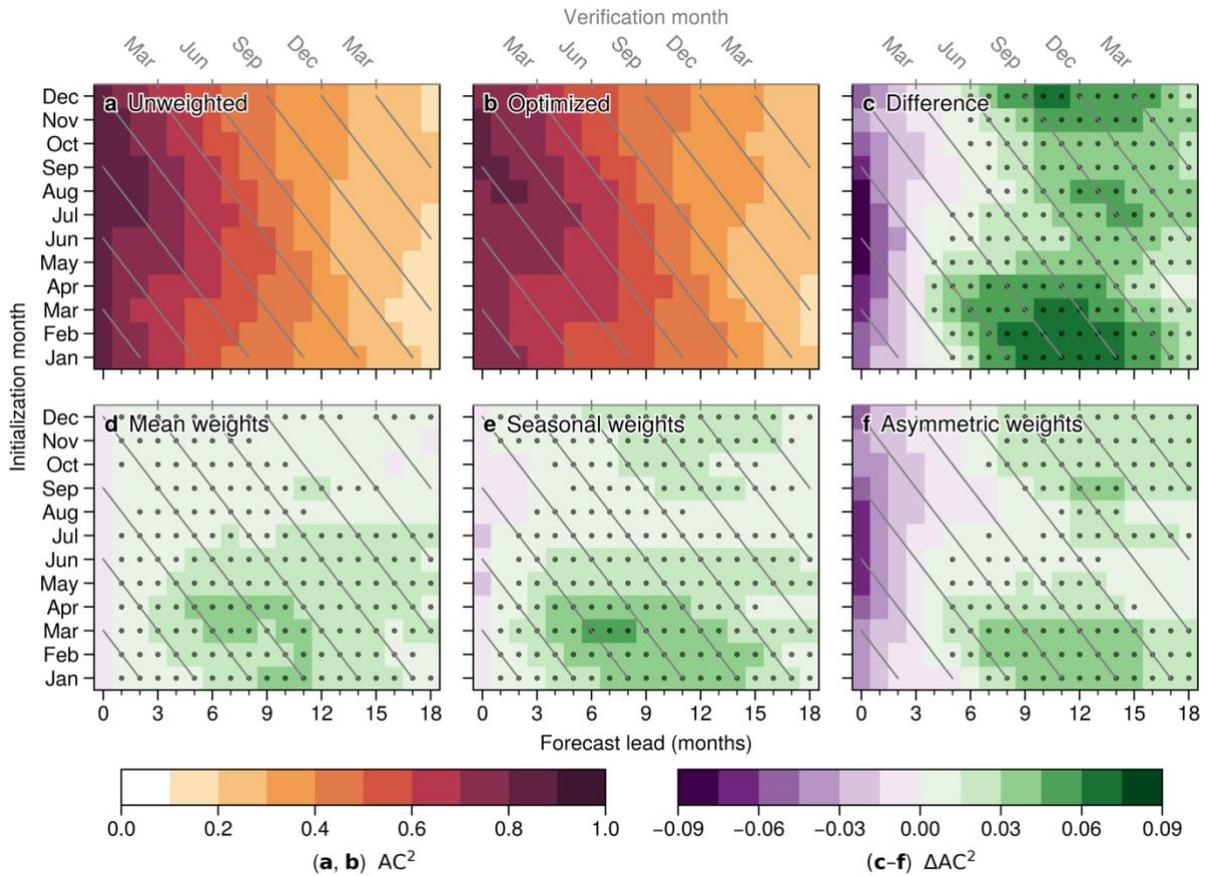

Fig. 4. The seasonality of squared anomaly correlation ($AC^2$) of SST averaged over the equatorial Pacific as a function of forecast lead for (a) the unweighted model-analog and (b) optimized model-analog. The difference in $AC^2$ between the unweighted model-analog and (c) the optimized model-analog, (d) the mean weights experiment, (e) the seasonal weights experiment, and (f) the asymmetric weights experiment. Stippling in (c–f) indicates statistically significant improvements according to the permutation test. The verification month is indicated by the gray diagonal lines.

Fig. 5 illustrates under which ENSO conditions prediction skill is improved, using January initialization with a 12-month lead time. It is evident that predictions of extreme events are significantly improved, for both El Niño and La Niña conditions (Fig. 5a), due to their large influences in the loss function. Conversely, predictions for ENSO neutral conditions (below 0.5 σ) show no discernible impacts on the median skill. The improvements in predicting extreme events diminish considerably when state-independent weights are applied, regardless of seasonality (Figs. 5b and 5c). Although improvements in extreme event predictions are still observed with the asymmetric weights experiment, they are not as significant (Fig. 5d). This indicates that the distribution and sign of input anomalies are crucial for estimating optimal weights in forecasting extremes.

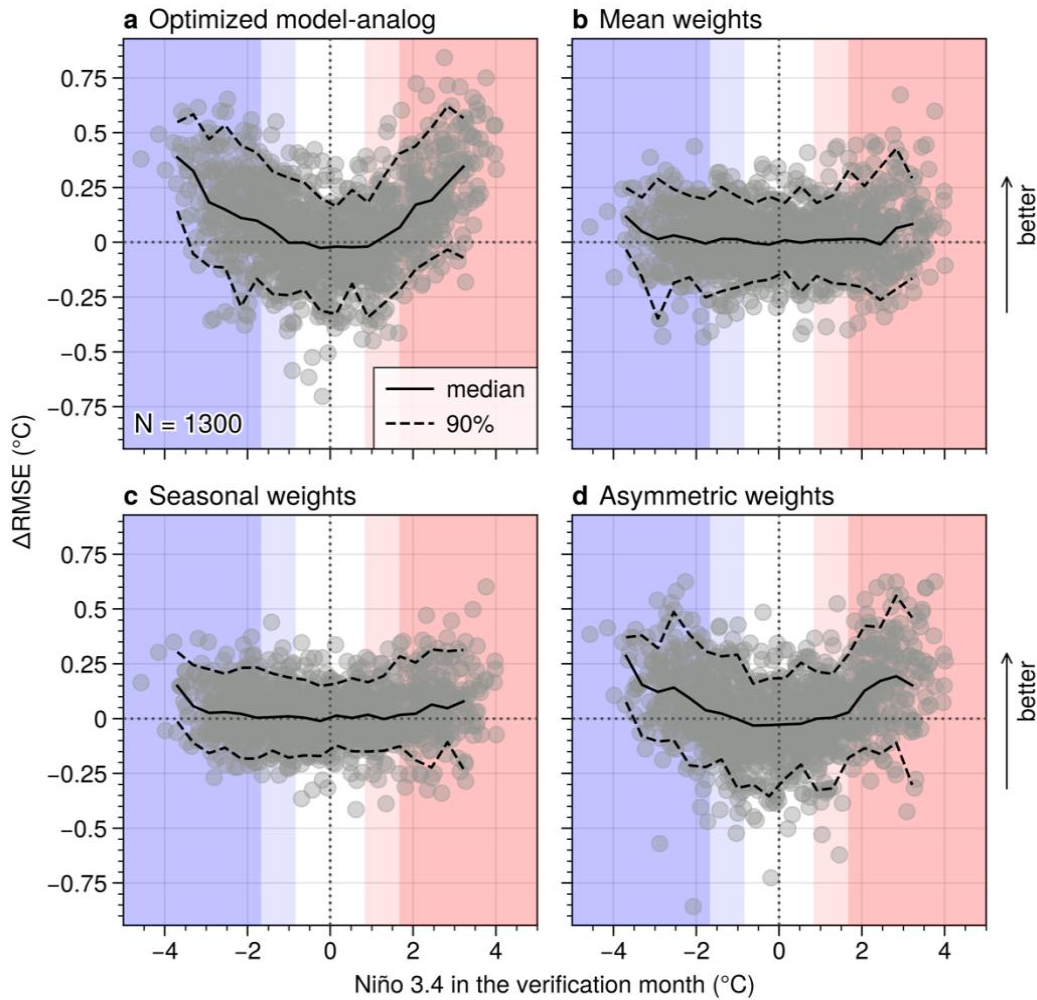

Fig. 5. Scatter plots of the RMSE reduction of SST over the equatorial Pacific and the Niño 3.4 index in the verification month for (a) the optimized model-analog, (d) mean weights, (e) seasonal weights, and (f) asymmetric weights experiments. The analysis focuses on 12-month forecasts initialized in January. Lighter pink/blue colors show values above 0.5 σ and darker pink/blue colors show values above 1 σ of the respective Niño 3.4 index. The median and 90% lines are estimated by binning samples according to the Niño 3.4 index.

Forecasting with analogs is by construction ensemble forecasting. The optimized model-analogs lead to similar probabilistic skill improvements, with reduced skill for shorter leads and enhanced skill for longer leads. This is seen in Fig. 6 which shows the all-month probabilistic forecast skill (CRPS) using 30 analog members. CRPS of 0.4°C is extended for more than 1 month in the all-month average. The improvements in CRPS are attributable to improvements in resolution (Fig. 6c), which may be anticipated given that the loss function is designed to penalize samples deviating significantly at forecast leads, resulting in narrower ensemble spreads. However, smaller ensemble spreads can deteriorate the reliability component, associated with the flatness of the rank histogram, as appears to have occurred in our results (Fig. 6b). The rank histogram is the frequency of the rank of the verification relative to sorted ensemble members. In the absence of ensemble variability, the rank histogram tends to exhibit a U-shaped distribution (Hamill 2001). Since ensemble reliability was not explicitly considered in the loss function, this stands as one of the caveats in this study.

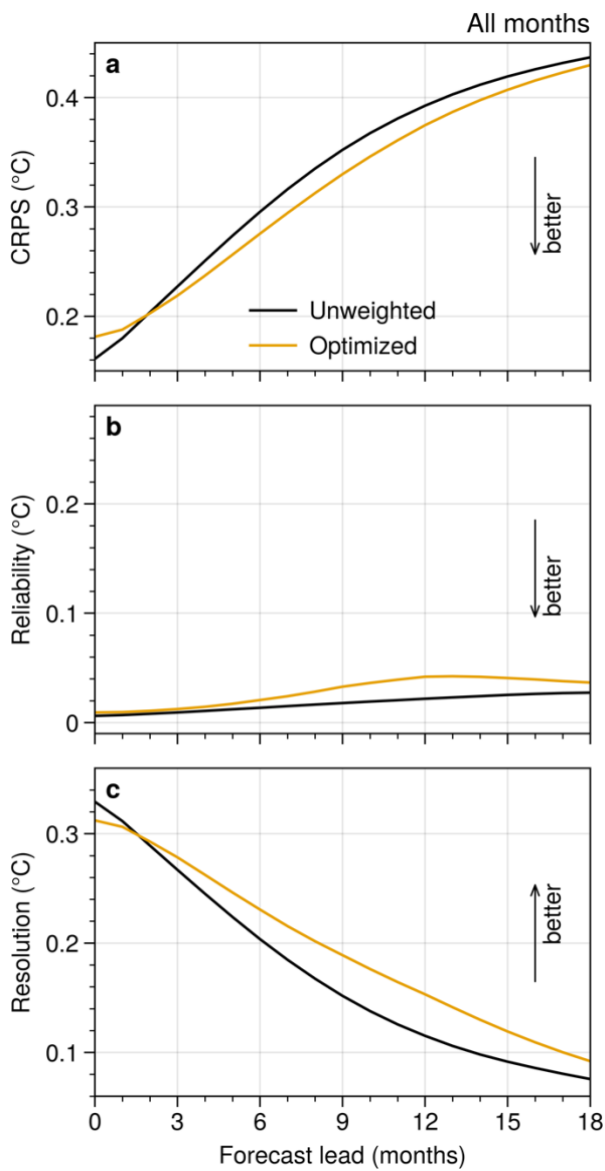

File generated with AMS Word template 2.0

Fig. 6. (a) Seasonally-averaged continuous ranked probability score (CRPS) of SST over the equatorial Pacific as a function of forecast lead by the unweighted and optimized model-analog methods. Similar to (a), but for (b) reliability and (c) resolution components of the CRPS.

Once model-analogs are identified, forecasting can be extended to any field available in the climate simulation. This is a distinct advantage in analog forecasting not achievable solely with neural networks, where predictors and predictands must be carefully chosen based on specific phenomena targeted by the model and the available computational resources. This approach may be particularly useful for precipitation forecasting where the signal-to-noise ratio tends to be low. Fig. 7 shows the improvements in 12-month precipitation forecasting using the optimized model-analog trained for equatorial Pacific SST. Precipitation forecasting is particularly improved in DJF (Fig. 7a), with significant improvements extending beyond the target region including the central subtropical Pacific, Maritime Continent, southwest Pacific east of Australia, southeastern US, northeastern Brazil, and north of Madagascar, potentially linked to ENSO teleconnections. Similarly, forecast skill in MAM is improved both within and outside the target region, albeit with smaller magnitudes (Fig. 7b). While precipitation forecast skill in JJA and SON also displays significant improvements, the impact is primarily confined within the target region (Figs. 7c and 7d). It is essential to highlight that, while not always statistically significant, positive impacts on precipitation forecasting are observed in most regions across all seasons (not shown). This suggests that improving the model-analog forecasts of tropical SST contributes positively to global precipitation forecasting.

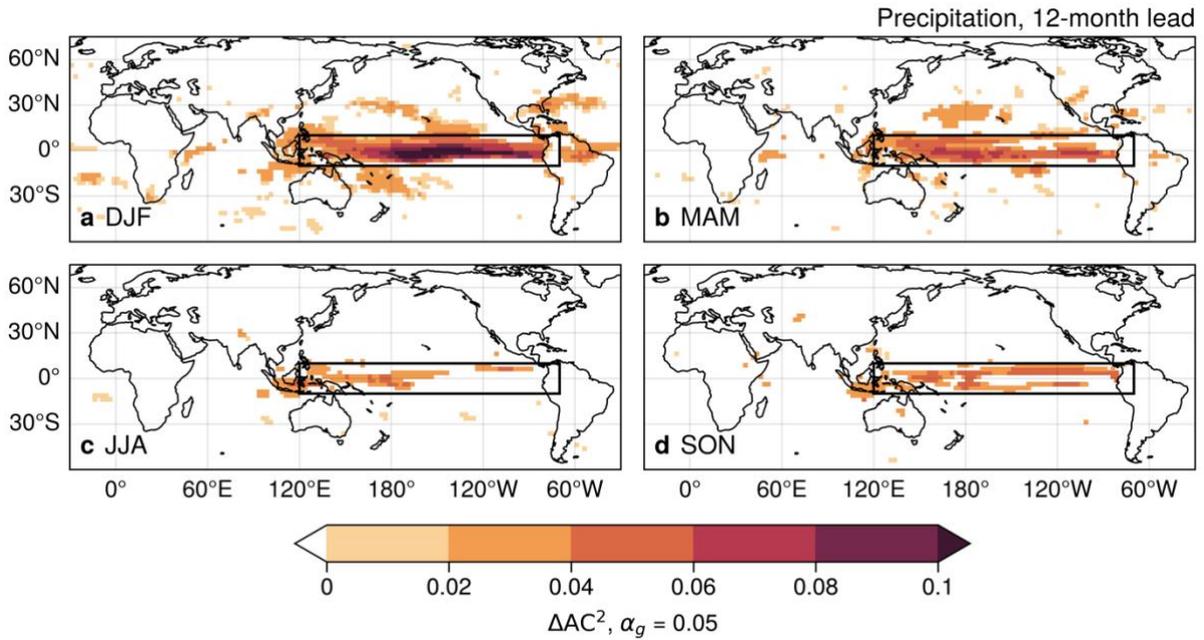

Fig. 7. Increase in squared anomaly correlation ($AC^2$) of 12-month lead precipitation by the optimized approach compared to the unweighted approach. The forecasts are initialized and verified for (a) DJF, (b) MAM, (c) JJA and (d) SON. Color shading indicates statistically significant improvements at the 5% level adjusted with the 5% false discovery rate.

## 4. Interpretable weights

*a. Spatial and seasonal variation of weights*

The neural network in the optimized model-analog approach produces interpretable weights whose state-dependence significantly impacts forecast skill (Figs. 4 and 5). These weights indicate sensitivity to initial error and should not be confused with precursors, which are early indicators of specific events. As in XAI methods, these weights do not provide causal relationships. Instead, they highlight the regions and variables where it is particularly important for the model-analogs to match the initial target anomalies, which will thereby most effectively constrain subsequent anomaly evolution through both physical processes and correlated or dependent features.

Fig. 8 illustrates the mean weights for four initialization months using the CESM2 test dataset. While the relative magnitudes of weights have a seasonal dependence, the weights are generally allocated to similar regions year-round. Notably, there are nonzero weights outside the target region (equatorial Pacific SST, indicated by the black box), although most of the weights are distributed within the tropics (30°S–30°N), suggesting that extratropical contributions are relatively small. These distributions of weights result in selecting analogs with poorer initial match (yet better subsequent trajectories) over the target region than unweighted model-analogs.

The distribution of weights among the three variables varies by calendar month, as shown in Fig. 9. From October to March, the weights are distributed relatively evenly between SST and SSH, with smaller weights for TAUX. April presents a deviation, with SST receiving the largest weights followed by SSH and TAUX. From May to September, the emphasis shifts, with TAUX receiving larger weights compared to SSH. Notably, TAUX receives the largest weights among all variables during June and July.

The spatial distributions of weights reveal connections to various physical processes associated with ENSO. In January (Fig. 8a) and April (Fig. 8d), SST receives weights that extend southwestward from the California coast toward the western equatorial Pacific, as well as over the eastern equatorial Pacific. This pattern closely resembles the characteristics of NPMM (Chiang and Vimont 2004; Amaya 2019), a robust predictor of ENSO conditions (Penland and Sardeshmukh 1995; Larson and Kirtman 2014; Vimont et al. 2014; Capotondi and Sardeshmukh 2015; Capotondi and Ricciardulli 2021). We find that largest weights in the NPMM region occur from April to June (Fig. 10a), which is also when the NPMM is typically strongest. Additionally, the SST weights in the subtropical southeastern Pacific resemble the pattern of the South Pacific Meridional Mode (SPMM) (Zhang et al. 2014), particularly evident in January (Fig. 8a) and October (Fig. 8j). The air-sea coupling associated with SPMM peaks in boreal winter (You and Furtado 2018), again consistent with when the SPMM weights are maximized (Fig. 10b). Regarding the July initialization (Fig. 8g), SST weights concentrate more over the eastern equatorial Pacific. This reflects the timing of ENSO events in boreal winter and their influences on subsequent seasons, which are the target leads of the July initialization.

SSH weights are consistently confined over the equatorial Pacific throughout the year, unlike SST (Figs. 8b, e, h, and k). Since SSH is dynamically linked to thermocline depth, this pattern likely relates to the recharge and discharge of upper-ocean heat content during the alternation of warm and cold ENSO phases (Jin 1997). In particular, a recharged state is conducive to the development of an El Niño, while a discharged state may likely lead to a La Niña. The equatorial weights can constrain the zonal tilt of the equatorial thermocline

File generated with AMS Word template 2.0

concurrent with the peak of ENSO, in addition to the recharge-discharge mode which is an important precursor of ENSO (Meinen and McPhaden 2000). Notably, these weights are particularly amplified in April (Fig. 10c). Equatorial Pacific upper-ocean heat content typically precedes Niño 3.4 SST by a quarter of the ENSO cycle (McPhaden 2003), equating to about 8–10 months in CESM2 (Capotondi et al. 2020). Given that ENSO events tend to peak in boreal winter, the peak of weights in April is consistent with these established temporal dynamics.

Winds play a crucial role in driving ENSO variability. TAUX weights tend to be largest in the western to central tropical Pacific throughout the year (Figs. 8c, f, i, and l), coinciding with the typical occurrence of stochastic wind forcing across the region. This stochastic forcing exhibits a broad spectrum ranging from subseasonal to interannual scales, with the lower frequency components exerting a greater influence on ENSO evolution (Roulston and Neelin 2000; Capotondi et al. 2018). During boreal summer, the absence of the interannual stochastic wind component can severely limit ENSO growth (Menkes et al. 2014), consistent with the peak magnitude of wind weights observed in June (Fig. 10d).

Although the target region lies within the tropical Pacific, allocation of weights to the Atlantic and Indian Ocean indicates the impact of tropical interbasin interactions (Cai et al. 2019; Wang 2019). Interestingly, over the Atlantic Ocean larger weights are distributed to SSH compared to SST (Fig. 9). Our result suggests that ocean memory (i.e., upper ocean heat content) may serve as a more reliable proxy for Atlantic influences compared to SST, which measures surface heat. In contrast, large SST weights are observed over the Indian Ocean in January and April (Figs. 8a and d), potentially linked to the Indian Ocean Dipole.

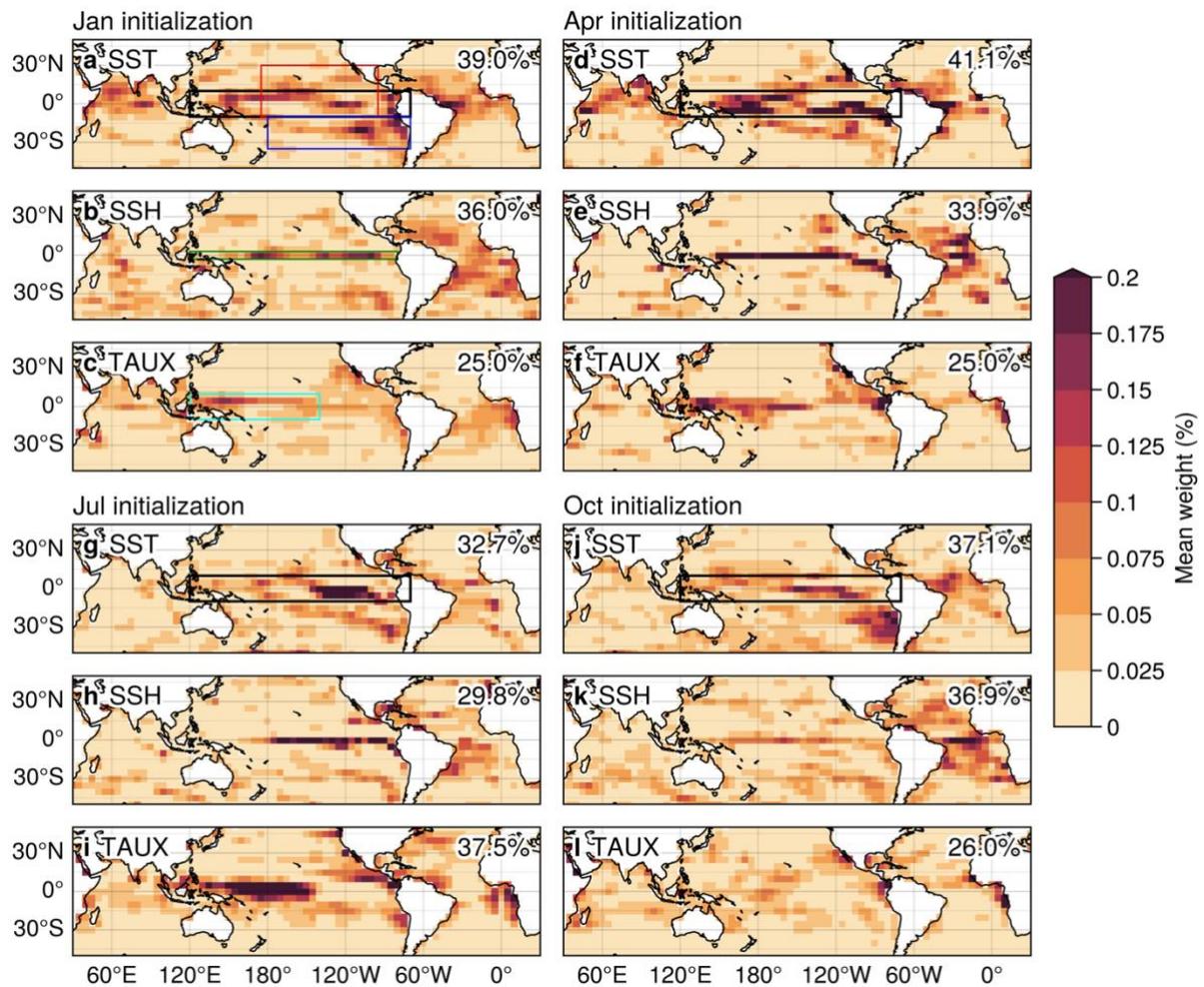

Fig. 8. Mean weights for (a–c) January, (d–f) April, (g–i) July, and (j–l) October initialization in the CESM2 test dataset. These weights improve the selection of analogs for forecasts with lead times of 6–18 months. Weights are unitless and scaled to ensure a sum of 100%. The sum of weights for each variable is displayed within each respective panel. Regions of interest, denoted by red (NPMM SST), blue (SPMM SST), green (equatorial Pacific SSH), and cyan (western to central tropical Pacific TAUX) boxes, are analyzed in Fig. 10.

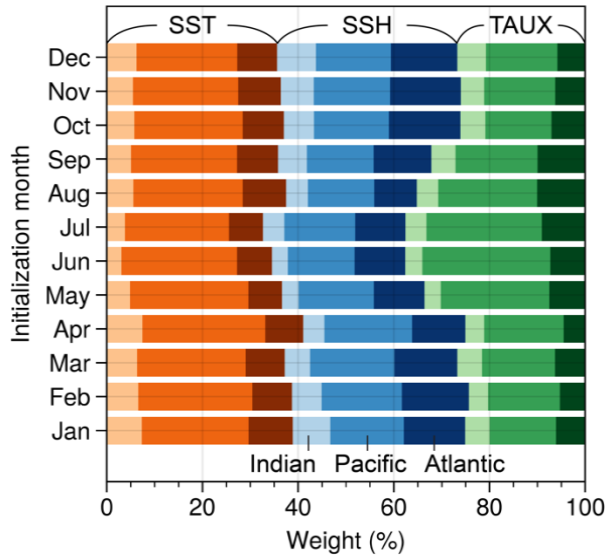

Fig. 9. Seasonal variation of mean weights in the CESM2 test dataset. Red, blue, and green represent the total weights for SST, SSH, and TAUX, respectively. The intensity of light, medium, and dark colors indicates the sum of weights over the Indian, Pacific, and Atlantic Oceans, respectively.

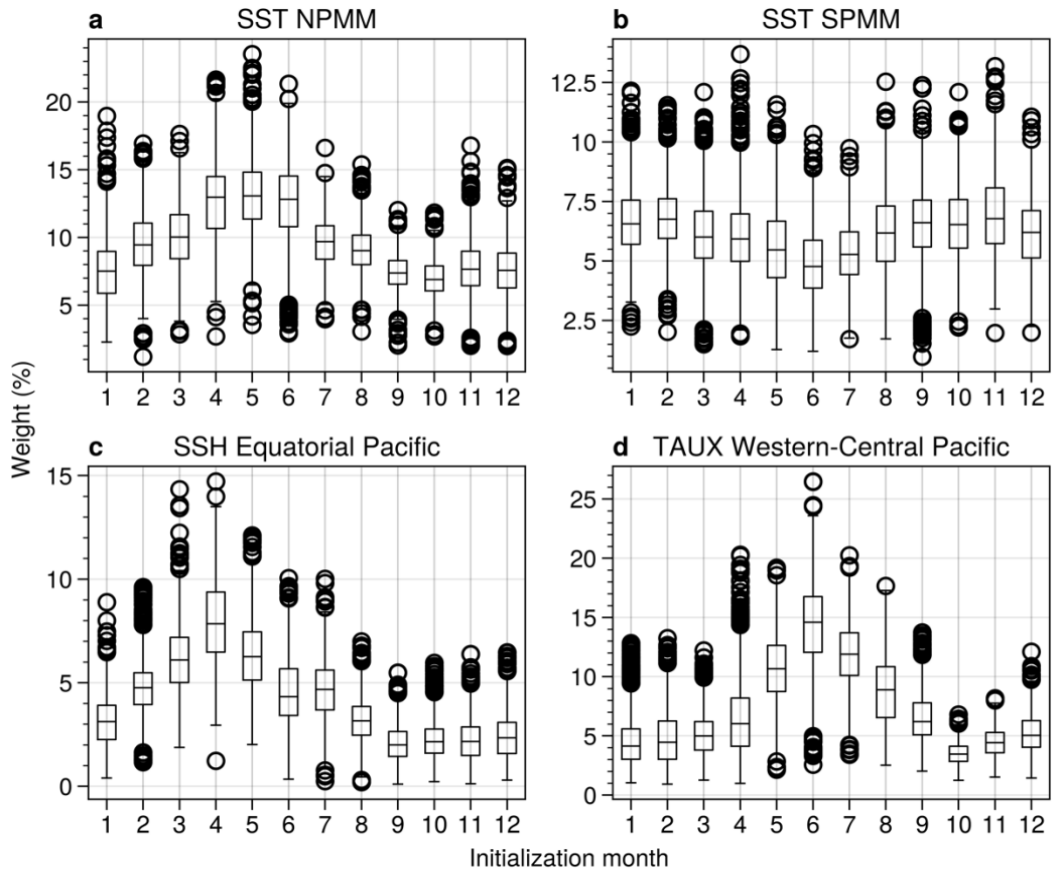

File generated with AMS Word template 2.0

Fig. 10. Seasonal variation of (a) SST weights over the NPMM region (10°S–30°N, 175°E–85°W), (b) SST weights over the SPMM region (35°S–10°S, 180°–70°W), (c) SSH weights over the equatorial Pacific (2.5°S–2.5°N, 120°E–80°W), and (d) TAUX weights over the western to central tropical Pacific (10°S–10°N, 120°E–140°W), as observed in the CESM2 test dataset. Box plots depict the minimum, maximum, median, first and third quantiles, and outliers.

*b. State-dependence and asymmetry in weights*

Since weights are state-dependent, we can analyze the asymmetry in sensitivity associated with El Niño and La Niña. Fig. 11 shows the difference in mean weights for events evolving to El Niño and La Niña, initialized in January, March, May, and July. Here, El Niño and La Niña events are defined by above and below ±0.5 σ of the Niño 3.4 index, respectively. A positive (negative) difference indicates that the prediction of El Niño (La Niña) is more sensitive to initial error in the specific region.

Generally, larger differences are observed for shorter lead times, as expected. The spatial distribution of sensitivity differences varies significantly with lead time. Pacific SST plays a crucial role in El Niño development during January to May (Figs. 11a, d, and g), while central equatorial Pacific SST is more important for La Niña development in July (Fig. 11j).

There is no consensus on the asymmetric predictability of ENSO associated with ocean heat content. Planton et al. (2018) found that a discharged warm water volume in the western equatorial Pacific is a better predictor of La Niña 13 month later, while Larson and Pegion (2020) suggest that the spread of ENSO phases is large when conditioned by the recharge-discharge. The difference in SSH weights in January indicates that the prediction of El Niño is more sensitive to errors in the equatorial Pacific heat content than La Niña (Fig. 11b), which contradicts Planton et al. (2018). From March to July, the difference in SSH weights exhibits a zonal dipole pattern in equatorial Pacific, where the central part is more important for El Niño and the eastern edge is more important for La Niña (Figs. 11e, h, and k).

In terms of tropical Pacific wind stress, weights are higher for La Niña in January (Fig. 11c). In May, the western tropical region is more important for La Niña, while the central tropical region is important for El Niño (Fig. 11i). By June and July, when wind stress weights become most significant, they are more important for El Niño (Fig. 11l), consistent

File generated with AMS Word template 2.0

with previous findings that El Niño is more influenced by zonal wind stress (Dommenget et al. 2013).

Overall, the asymmetry in sensitivity varies significantly with lead time, which may partially explain the challenges in attributing ENSO asymmetry to various nonlinear processes (An et al. 2020).

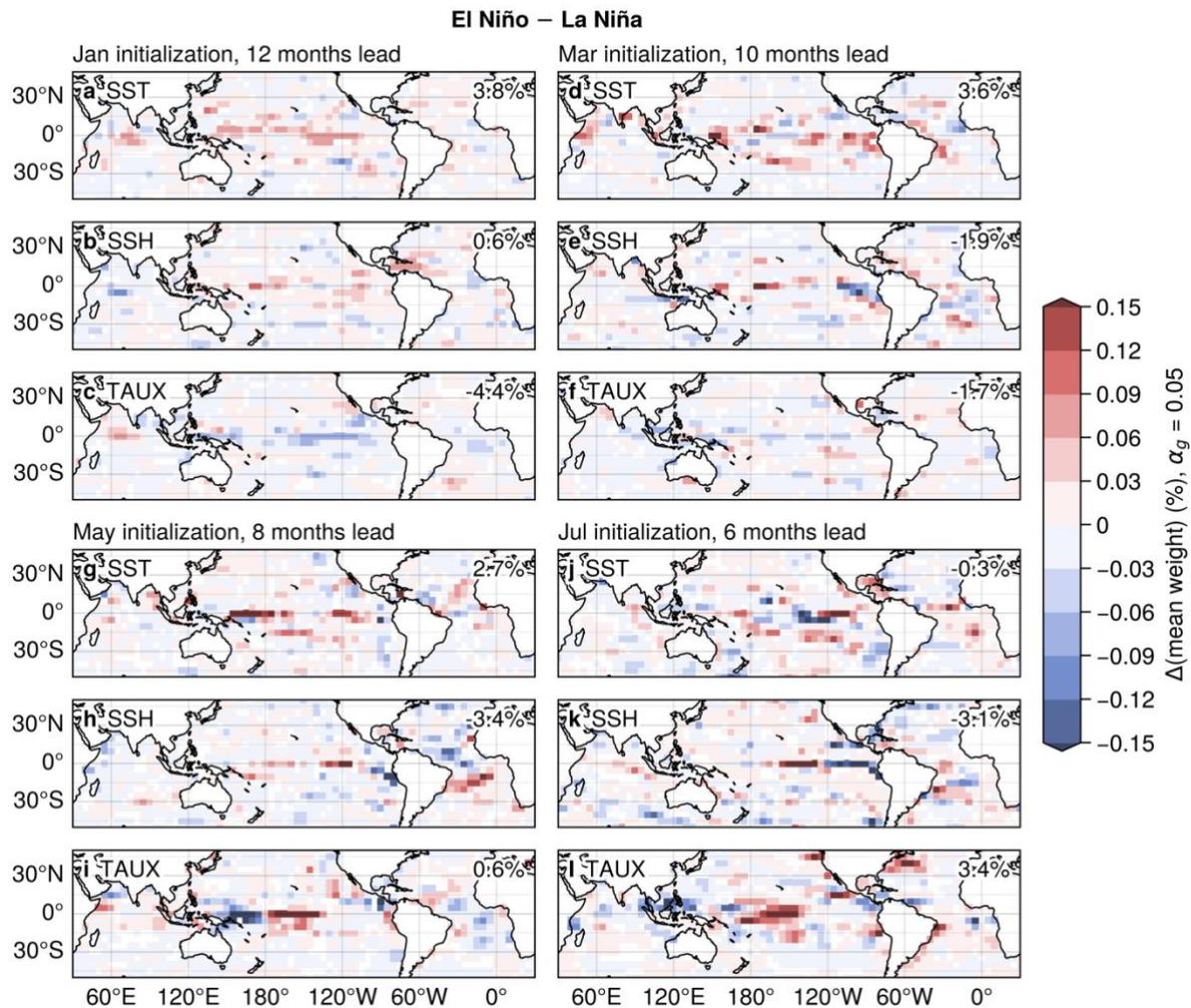

Fig. 11. Differences in mean weights for events evolving into El Niño and La Niña conditions by January of the following year, with initializations in (a–c) January, (d–f) March, (g–i) May, and (j–l) July. Color shading indicates statistically significant differences at the 5% level adjusted with the 5% false discovery rate. Red indicates larger weights for El Niño prediction and blue indicates larger weights for La Niña prediction.

# 5. Application to observations

We next apply the developed optimized model-analog approach to make real-world hindcasts by finding optimized model-analogs for initial anomalies drawn from the ORAS5 reanalysis dataset, using the same network trained with CESM2. Recall that we do not use any observations to train the optimized model-analog technique, nor do we employ transfer learning for these hindcasts. Fig. 12 shows the seasonal variation of hindcast skill during 1987–2020. The original (unweighted) model-analog shows lower skill than the perfect-model skill (Fig. 4a) with a spring predictability barrier where skill sharply declines around March (Fig. 12a). The impact of the optimized approach varies across initialization months (Fig. 12c), in a manner that is broadly similar to its impact upon perfect model skill (Fig. 4c). However, although positive effects are observed in many initialization months, forecasts initialized in Aug–Oct display a decrease in skill. Statistically significant improvements are observed in boreal fall forecasts initialized in May and June, as well as in year 2 spring forecasts initialized in boreal winter.

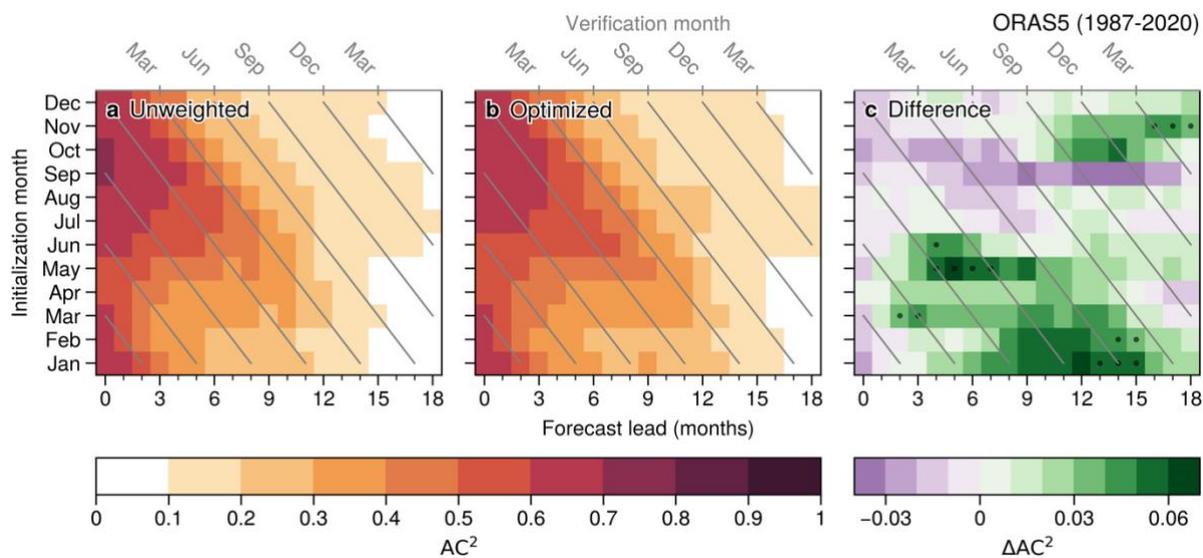

Fig. 12. Similar to Figs. 4a–c, but for hindcast initialized during 1987–2020 using ORAS5.

Fig. 13 shows the ENSO conditions under which prediction skill is improved for observationally-based hindcasts. Although the sample size is small, predictions of extreme events are also improved in the hindcasts. Apart from the La Niña event in 1996, the optimized approach reduces forecast error for all extreme events above 1 σ (darker shading). However, issues with model errors could also play a role. In Fig. 5a, the optimized approach significantly improves extreme event forecasts, particularly those characterized by Niño 3.4 values much higher than historically observed values. This result suggests that the neural network may be learning some information with limited relevance to the real world.

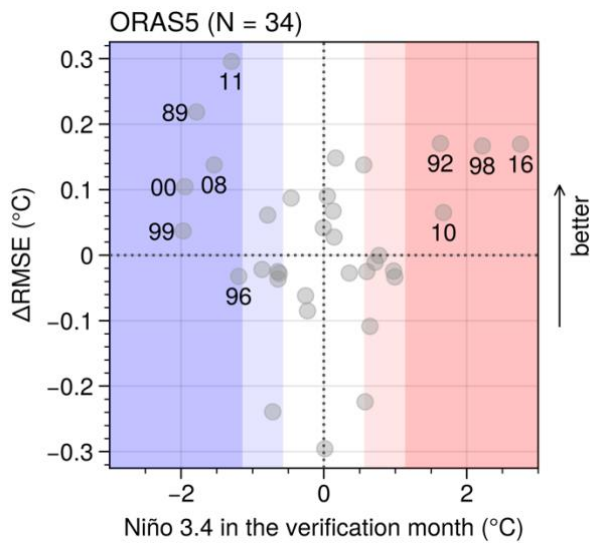

Fig. 13. Similar to Fig. 5a, but for hindcast initialized during 1987–2020 using ORAS5. The last two digits of verification years are displayed for extreme events.

## 6. Conclusion

In this study, we introduce an interpretable-by-design machine learning approach called the optimized model-analog method, which uses deep learning to generate state-dependent weights for selecting analog members. This approach offers two aspects of interpretability. First, the estimated weights, which serve as the explanations by the network, highlight regions that are particularly sensitive to initial error. Unlike post-hoc explanations provided by XAI, this reasoning process is inherently built into the network. Second, model-analog forecasts are derived from physically-based climate simulations. The cause-and-effect relationships within these models are based on established physical laws and principles. This contrasts with the black-box nature of traditional neural networks, where individual neuron interactions are difficult to interpret. We demonstrate that our two-step approach can enhance the potential of model-analog forecasting and yields forecast skills comparable to those of a standalone neural network approach.

File generated with AMS Word template 2.0

The application to ENSO forecasting shows significant improvements compared to the original (unweighted) model-analog method in perfect model skill at 6–18 months leads. The most significant improvements are observed in the central equatorial Pacific region and in predicting extreme events due to the large SST variability. We further show that state-dependent weights are crucial for these improvements by comparing them with the state-independent weights and asymmetric weights experiments. Once optimized model-analogs are identified based on weighted distances, their subsequent time evolution can be analyzed in any fields available in the original climate simulation dataset. We demonstrate that improving equatorial Pacific SST forecasts also results in improving precipitation forecasting beyond the target region.

The hybrid approach predicts weights linked to various known seasonally varying physical processes. Specifically, SST weights exhibit patterns similar to NPMM peaking in boreal spring and SPMM peaking in boreal winter. SSH weights are concentrated over the equatorial Pacific, likely capturing states linked to the recharge-discharge of warm water volume associated with ENSO oscillatory behavior. TAUX weights are large in regions where stochastic wind forcing typically occurs, with a peak in boreal summer. Furthermore, some weights are distributed over the Atlantic and Indian Ocean, indicating the influence of the tropical interbasin interactions. The asymmetry in ENSO forecasting is also observed: El Niño forecasts are more sensitive to initial error in tropical Pacific SST in boreal winter, while La Niña forecasts are more sensitive to initial error in tropical Pacific TAUX in boreal summer. These weights are generated by the neural network method used, implying that it is straightforward to integrate superior deep learning algorithms for improved weight quantification.

We additionally show improvements in hindcast applications using ORAS5 across many initialization months and extreme events, although certain initialization months exhibit a reduction in forecast skill. Several factors contribute to the differences between hindcast and perfect-model results. Climate models inherently possess systematic errors, such as the excessive westward extension of the SST anomalies associated with ENSO (Bellenger et al. 2014), which is also evident in the CESM2 model (Capotondi et al. 2020) and in all seasonal climate model forecasts (Newman and Sardeshmukh 2017; Beverley et al. 2023). If the neural network learns a model attractor that is significantly different from reality, it can deteriorate skill. A potential solution to mitigate model biases involves employing multiple climate models, as demonstrated in model-analog studies (Ding et al. 2018, 2019; Lou et al. 2023), and machine learning studies (Ham et al. 2019; Zhou and Zhang 2023). Transfer learning may also alleviate biases, although with limitations due to sample size and the effects of climate change. Additional reasons for less significant results include a limited sample size, uncertainty in the fair-sliding anomaly calculation method, and uncertainty in the reanalysis dataset used both to choose initial model-analogs and to verify the subsequent hindcasts. Future work should address these challenges by mitigating the effects of model biases, potentially through the incorporation of multiple climate models and leveraging transfer learning techniques, and by developing hindcasts based on multiple different reanalysis datasets.

Our approach mirrors the principles of adjoint sensitivity, where a linearized model is used to assess the sensitivity of a specific aspect of the final forecast to initial conditions (Errico 1997). Recently, Vonich and Hakim (2024) expanded on this concept by using neural networks to estimate optimal initial conditions by iteratively minimizing forecast errors through backpropagation and gradient descent. Additionally, our method can be viewed as a nonlinear and flow-dependent extension of singular vectors (Diaconescu and Laprise 2012) or optimal perturbations (Penland and Sardeshmukh 1995). These methods identify perturbations with maximum growth under a specific norm over a finite time interval. Despite the conceptual similarities, our approach stands out by not requiring a predefined target once trained when forecasting from a given initial condition.

There are many possible applications of this approach. It can be used for different climate phenomena across various regions, such as regional temperature and precipitation. This has been challenging with the unweighted model-analog because the selection of input variables and input regions must be made for each target, which could be subjective. The optimized model-analog approach addresses this issue by optimizing the focus (i.e., weights) in the input space using neural networks.

Another application is evaluating the regional and variable contributions to forecasting skill, including the assessment of interactions between the tropical basins. Broadly, two approaches can be considered: 1) training neural networks with restricted regions/variables, and 2) modifying (i.e., zeroing) predicted weights of certain regions/variables. The first approach may yield results that are difficult to interpret due to correlations between used and unused features. On the other hand, the latter approach involves post-modification after model training and selects analogs without constraining a part of the input. This approach could provide interesting insights into quantifying the contribution of a specific feature by allowing error growth from that feature.

*Acknowledgments.*

This work was supported by the Famine Early Warning Systems Network and the NOAA Physical Sciences Laboratory. Antonietta Capotondi was supported by DOE Award No. DE-SC0023228. Jakob Schlör was supported by EXC number 2064/1 – Project number 390727645 and the International Max Planck Research School for Intelligent Systems (IMPRS-IS). The authors thank Tim Smith, Jannik Thuemmel, Elizabeth Barnes, and three anonymous reviewers for comments that improved this work.

*Data Availability Statement.*

The CESM2-LE dataset is available from The National Center for Atmospheric Research (https://doi.org/10.26024/kgmp-c556). The ORAS5 dataset is available from the European Centre for Medium-Range Weather Forecasts (https://doi.org/10.24381/cds.67e8eeb7). The optimized model-analog codes are publicly available on GitHub (https://github.com/kinyatoride/DLMA).

Supporting Information for

# Using Deep Learning to Identify Initial Error Sensitivity for Interpretable ENSO Forecasts

Kinya Toride,[a,b] Matthew Newman,[a] Andrew Hoell,[a] Antonietta Capotondi,[a,b] Jakob Schlör,[c] Dillon J. Amaya,[a]

[a] *Physical Sciences Laboratory, National Oceanic and Atmospheric Administration, Boulder, Colorado*

[b] *Cooperative Institute for Research in Environmental Sciences, University of Colorado Boulder, Boulder, Colorado*

[c] *European Centre for Medium Range Weather Forecasts (ECMWF), Reading, UK*

*Corresponding author*: Kinya Toride, kinya.toride@noaa.gov

**Contents of this file**

Texts S1 to S3

Figures S1 to S3

**Text S1 Network size**

The complexity of a model, often indicated by the number of parameters, plays an important role in machine learning studies. Although the trend in the field leans towards more complex models with advanced skill, it is equally important to explore the potential gains achievable with simpler models, especially for those with resource constraints. As described in the Methods section, the network size is controlled by two key hyperparameters: depth and initial channel size. We employ a depth of 4 and an initial channel size of 256 in this study (referred to as 4-256), resulting in 123 million trainable parameters. This is determined through hyperparameter tuning and training cost considerations.

Either reducing the depth by 1 or halving the initial channel size decreases the number of parameters by a factor of four. We found that reducing the depth degrades model performance more than reducing the initial channel size. This may be due to the reduction in the receptive field size, which represents the region in the input space influencing an output in a single grid, associated with decreasing depth. Since forecasting ENSO requires capturing large-scale teleconnections as illustrated in the estimated weights (Fig. 9 in the manuscript), maintaining a deep network is imperative. Although it is tempting to have a deeper network, the current input size limits the depth to 4.

Therefore, we conduct a sensitivity analysis by varying the initial channel size. Fig. S1a shows the reduction in RMSE on the validation dataset for different network sizes. As the network size increases, the skill improvement follows an asymptotic trend. Statistical tests reveal no significant difference between the 4-256 model and the 4-64 model, which has 16 times fewer parameters. Yet, a significant difference is observed between the 4-512 and 4-64 models (not shown). Hence, one needs to consider the trade-off between computational costs and model performance.

The training duration for the 4-256 model is approximately 30 minutes and 1 hour with a single NVIDIA A100 and A6000 GPU, respectively (Fig. S1b). While the training time decreases with a smaller model, the difference diminishes for models with an initial channel size smaller than 128. This is due to the sorting of samples in the library. With smaller networks, sorting time dominates, while larger networks exponentially increase training time. It is essential to note that actual training time and sensitivity to network size may vary depending on the system used.

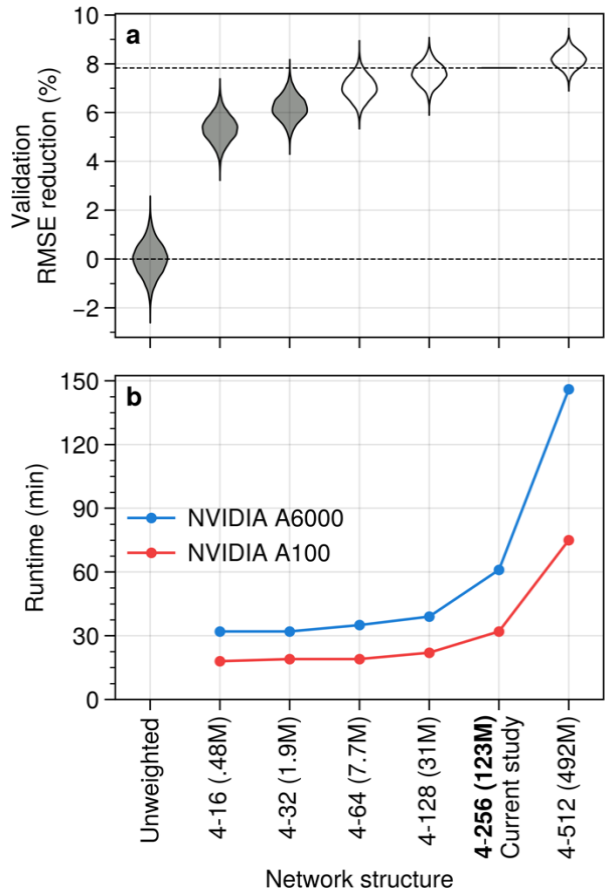

Fig. S1. (a) RMSE reduction (%) of 12-month lead SST over the equatorial Pacific in the validation dataset for different network structures. The network structure is denoted by depth-(initial channel size) with parameter counts in parentheses. Violin plots illustrate the null distribution estimated through permutation with the 4-256 model results. Gray shading indicates values are significantly different at a 5% level. (b) Approximate time taken to train U-Net models for 60 epochs using a single NVIDIA A6000 or A100 GPU in this study.

**Text S2 Unweighted model-analog**

This section presents the sensitivity of unweighted model-analog results to some parameters. Fig. S2a shows a skill comparison among different input regions and variables. The highest skill is achieved with SST and SSH over the tropics (30°S–30°N), as used in Lou et al. (2023). Expanding the input domain to the extratropics and including TAUX lead to a degradation in skill. Although the optimized model-analog approach assigns weights to the three variables over 50°S–50°N, we choose the one with SST and SSH over the tropics to avoid underestimating the skill of the unweighted approach.

Fig. S2b shows the sensitivity to analog member size. RMSE clearly worsens with a member size of fewer than 10. We select a member size of 30, which minimizes RMSE at lead times of 6–12 months.

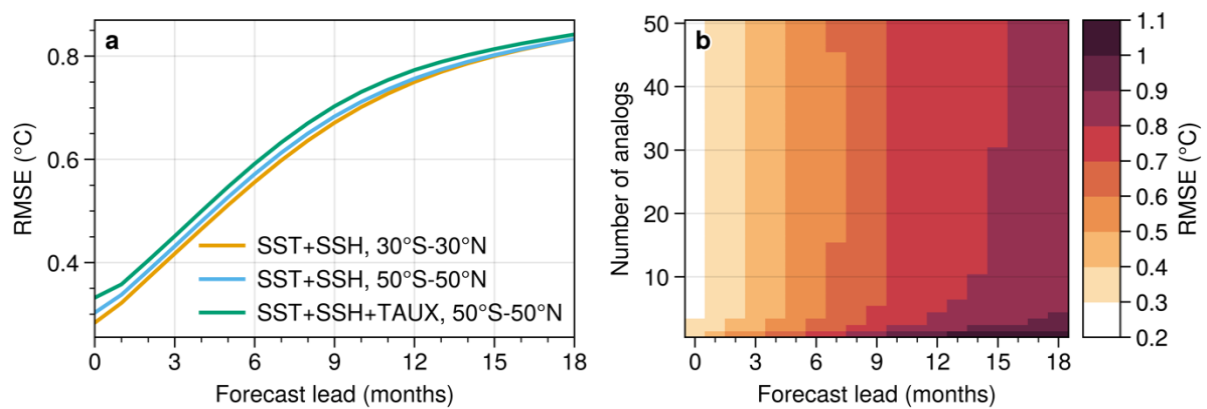

Fig. S2. (a) RMSE of equatorial Pacific SST as a function of forecast lead on the test dataset. Three unweighted model-analog approaches with different inputs are evaluated. (b) RMSE of equatorial Pacific SST as a function of forecast lead and analog member size.

File generated with AMS Word template 2.0

**Text S3 Lead time dependence**

Fig. S3 shows a comparison of RMSE reduction using different forecast errors in the loss function. The model is trained with MSE at a specific lead time (3, 6, 9, or 12 months) in addition to using averaged MSE over 3, 6, 9, and 12 months leads. Note that the learning rate is fine-tuned independently. While the training results with a lead time of 3 months exhibit significantly different behavior, other results display more similarity. This tendency is also observed in the estimated weights, where the 3-month lead results focus more on the tropical Pacific (not shown). Among longer leads, the 6-month lead results yield the highest skill, especially for shorter leads. The results with the averaged MSE are slightly worse around 6-month lead but generally comparable to the 6-month lead results. Considering the potential dependency of training results on the initial month, we use the averaged MSE in this study.

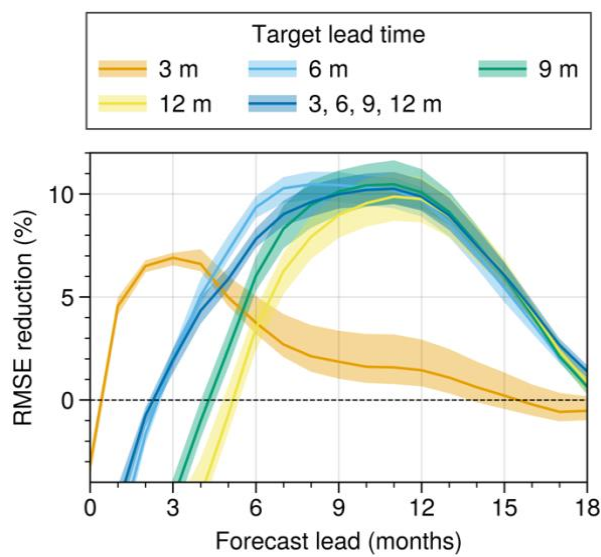

Fig. S3. RMSE reduction (%) of equatorial Pacific SST as a function of forecast lead for January initialization using the test dataset. The optimized model-analog is trained for various lead times. Shading shows the spread due to random initialization of network parameters.